\documentstyle[12pt]{article} 
\advance\textheight by \topskip 
\begin{document} 
\input epsf

\thispagestyle{empty}
\begin{flushright} 
{SU-ITP-96-03} \\ hep-th/9601083 \\ January 16, 1996\\
\end{flushright}
\vskip 3cm
\begin{center} 
{\Large\bf Relaxing the Cosmological Moduli Problem} \vskip 1.1cm {\bf
Andrei Linde} \\ \vskip
0.5cm Department of Physics, Stanford University, Stanford CA
94305-4060, USA
\end{center}

\vskip 1 cm {\centerline{\large ABSTRACT}}
\begin{quotation} \vskip -0.3cm
Typically the moduli fields acquire mass $m_\phi^2  = \pm C^2 H^2$ in the early universe, which shifts the position of the minimum of their effective potential and leads to an excessively large energy density of the oscillating moduli fields at the later  stages of the evolution of the universe. This constitutes the cosmological moduli problem, or Polonyi field problem. We show that the cosmological moduli problem can be solved or at least significantly relaxed  in the theories in which   $C \gg 1$, as well as in some models with   $C \ll 1$.

\end{quotation}

\newpage

%%%%%%%%%%%%%%%%%%%%%%%%%%%%%%%%%%%%%%%%%%%%%%%%%%%%%%%%%%%%%%%
 
String moduli  couple
to standard model fields only through Planck scale suppressed
interactions. Their effective potential is
exactly flat  in perturbation theory perturbatively in the supersymmetric limit, but it may become curved due to nonperturbative effects or because of supersymmetry breaking. If these fields originally were far from the minimum of their effective potential, the energy of their oscillations decrease in an expanding universe in the same way as the energy density of nonrelativistic matter, $\rho_m \sim a^{-3}(t)$.  Meanwhile energy density of relativistic plasma decreases as $a^{-4}$. Therefore the relative contribution of   moduli to the energy density of the universe may soon become quite significant. They are expected to decay after the stage of nucleosynthesis, violating the standard nucleosynthesis predictions, unless the initial amplitude of the moduli oscillations $\phi_0$ is sufficiently small.  The constraint  on $\phi_0$ depends on details of the theory. The most stringent constraint appears because of the photodissociation and photoproduction of light elements by the decay products, $\phi_0 {\ \lower-1.2pt\vbox{\hbox{\rlap{$<$}\lower5pt\vbox{\hbox{$\sim$}}}}\ } 10^{-10} M_p$, see   \cite{constraint,rt} and references therein.  However, if one makes an assumption that moduli  decay only to the particles in the hidden sector, the constraint becomes less stringent, $\phi_0 {\ \lower-1.2pt\vbox{\hbox{\rlap{$<$}\lower5pt\vbox{\hbox{$\sim$}}}}\ } 10^{-7} M_p$. For greater values of $\phi_0$   the energy density of the oscillating field dominates the energy density of the universe at the epoch of  nucleosynthesis, which leads to    a significant overproduction of $^4He$ (i.e. to the absence of hydrogen).    Meanwhile one would expect the initial amplitude of oscillations  $\phi_0$ to be of the same order as $M_p$. (Here we use stringy normalization for the Planck mass, $M_p = {1\over \sqrt{8\pi G}} \sim 2\times 10^{18}$ GeV.) This is the essence of the cosmological moduli problem, which is  the string
version \cite{bkn,rt} of the Polonyi problem \cite{polonyi}.

There were   many suggestions how to solve this problem. For example, it was suggested that  the moduli fields slowly slide down to the minimum of their effective potential during inflation, and do not oscillate there anymore \cite{NS}. This regime would be possible even for very light moduli   if inflation is long enough. Moreover, according to \cite{dinefisch}, moduli fields typically acquire mass $m_\phi \sim H$ during inflation. Thus, their effective mass during inflation was not that small, and they could roll down to their minimum even if inflation was not very long \cite{NS,Dvali}. However, as was argued by Goncharov, Linde and Vysotsky  \cite{GLV}, this does not solve the problem since typically the minimum of the effective potential during inflation does not coincide  with the minimum of the effective potential at the present time with an accuracy $10^{-7} M_p - 10^{-10} M_p$. Recently this problem was investigated by Dine, Randall and Thomas \cite{adshort}, who have argued that the positions of the two minima may in fact coincide if one invokes some additional symmetries. If the mass of the moduli fields is very large (which may happen in certain models, see e.g. \cite{ENQ}), then they decay very early and do not pose any problems. A more general solution would be to have an additional stage of inflation which would dilute the energy of the oscillating moduli fields \cite{rt}. The most elegant realization of this idea is the ``thermal inflation'' scenario suggested by Lyth and Stewart \cite{LS}. It appears that in many models where scalar potentials have flat directions a secondary stage of inflation may indeed take place 
when the temperature becomes sufficiently small. This stage is short, but  it may be long enough to resolve the cosmological moduli problem. A similar (or maybe even somewhat longer) stage of ``nonthermal'' inflation may occur due to nonthermal phase transitions after reheating \cite{KLSphase}.

Thus, it may happen that after all the cosmological moduli problem may be not too severe. However, since ``thermal'' (or ``nonthermal'') inflation is very short,    it  solves the cosmological moduli problem only   under some optimistic assumptions about parameters of the models.
Therefore  it would be nice to have an additional mechanism which would help us to solve or at least to somewhat relax this problem.

To investigate this possibility one should note that there are many mechanisms giving the contributions $O(H^2)$ to $m_\phi^2$.   The sign and the magnitude of the sum of all these contributions is not well known. At the classical level, one would expect that in the early universe $m_\phi^2 \sim   H^2$, but this expectation can be strongly altered by quantum corrections and by nonrenormalizable terms in the effective Lagrangian  \cite{DRT}. For example, radiative corrections may lead to the terms in the effective Lagrangian of the form
\begin{equation}
\delta L =
 \pm{ C^2\over M_p^2} \int d^4 \theta ~ \chi^{\dagger} \chi \phi^{\dagger} \phi \ , 
\label{kahlercoupling}
\end{equation}
where $\chi$ is some field which dominates the energy density of
the universe.
At argued in \cite{DRT}, the existence of such operators is guaranteed
in the presence of Yukawa couplings since they are necessary
counterterms for operators
generated by loop diagrams. But this means that the coefficient $C^2$  in front of this term {\it a priori} can take any value; there is no reason to expect that $C^2$ is particularly small, or that $C^2 \sim 1$.
If $\chi$ dominates the energy density, then
$\rho \simeq \langle \int d^4 \theta \chi^{\dagger} \chi \rangle$.
The interaction
(\ref{kahlercoupling}) therefore gives a contribution $\Delta m^2_\phi =\pm C^2{ \rho \over  M_p^2}$ to the  effective mass of the field $\phi$. During inflation $\rho = 3 H^2 M_p^2$, i.e. $\Delta m^2_\phi = \pm {3 C^2 H^2}$.  Thus, we see that this contribution to the moduli mass squared is proportional to $H^2$, but the absolute value and the sign of the coefficient $\pm C^2$ is unknown since it is determined by the counterterms which appear in a nonrenormalizable theory. In general, one may    add  such counterterms with   any  coefficient $\pm C^2$.

The main observation which we are going to make is the following. The standard formulation of the moduli problem as we know it pertains only to the case $m_\phi^2 \sim \pm C^2  H^2$ with $C  \sim 1$. Meanwhile, under certain conditions  the moduli problem can be either completely solved or at least considerably relaxed   both for $C^2  \ll 1$ and for $C^2  \gg 1$.

We will begin with the discussion of the possibility $C^2 \ll 1$, which seems less natural since it requires some fine  tuning. In this case the moduli masses during inflation remain very small, and the motion of the field $\phi$ towards the minimum of its effective potential will be very slow. Due to quantum fluctuations during the early stages of inflation, the field $\phi$ takes {\it all } its possible values in different parts of the universe. Since these values do not change much at the   stage  with $C^2H^2 > m_\phi^2$, at the end of this stage the universe will consist of exponentially large domains with all or almost all possible values of the field $\phi$. In those domains where this field will be   displaced from the minimum of its   effective potential at small $H$ by more than $10^{-7} M_p$,  the energy of its oscillations will be very large. In such domains the density of the universe, and the speed of its expansion at the time of nucleosynthesis (which is determined by the value of temperature) would be much greater than in the part where we live. As a result, neutrons would not have enough time to decay, and all hydrogen would be conversed to helium. Therefore there are no hydrogen-burning stars in such parts of the universe, and we would be unable to live there. We would be able to live only in those (exponentially large)  parts of the universe where the initial amplitude of oscillations of Polonyi fields is small enough not to disturb nucleosynthesis. 

Note, that this does not explain why should the field $\phi_0$ be smaller than $10^{-10} M_p$, since photodissociation and photoproduction of light elements after the moduli decay does not seem to lead to any   problems for  the emergence of life for $\phi_0<10^{-7} M_p$. Therefore our proposal solves the problem only if $\phi_0<10^{-7} M_p$ is the most stringent constraint on $\phi_0$. This is the case for the moduli which decay   only to the particles in the hidden sector. Another potential problem with this explanation is that quantum fluctuations $\delta\phi \sim H/2\pi$ of the moduli field produced each time $H^{-1}$ during inflation lead to isothermal density perturbations ${\delta\rho\over\rho} \sim {\delta \phi\over\phi}$ \cite{MyBook}. These perturbations should not exceed   $10^{-5}$ to avoid conflict with the COBE data. This leads to the condition  ${H\over 2\pi\phi} {\ \lower-1.2pt\vbox{\hbox{\rlap{$<$}\lower5pt\vbox{\hbox{$\sim$}}}}\ } 10^{-4}$. For $\phi \sim 10^{-7} M_p$ this implies that $H {\ \lower-1.2pt\vbox{\hbox{\rlap{$<$}\lower5pt\vbox{\hbox{$\sim$}}}}\ } 10^8$ GeV.  In simplest models of chaotic inflation this condition  is not satisfied. However, it can be satisfied in hybrid inflation models \cite{hybrid}.  

Now let us consider a more interesting possibility, $C^2 \gg 1$. The standard assumption of all recent works on the cosmological moduli problem is the following. In the early universe the moduli field stays in the position corresponding to the minimum of the effective potential determined by the corrections $\sim H^2\phi^2$. When the Hubble parameter   becomes smaller than $m_\phi$,   the field $\phi$ rolls toward its present value {\it and oscillates about it with the amplitude $\phi_0$ approximately equal to the distance between the minima of the effective potential at large $H$ and at small $H$.} This last assumption seemed so natural that nobody actually verified it.

Meanwhile for $C^2 \gg 1$ this assumption is incorrect. Indeed, for $C^2 \gg 1$, $H {\ \lower-1.2pt\vbox{\hbox{\rlap{$>$}\lower5pt\vbox{\hbox{$\sim$}}}}\ } m_\phi$,  the effective potential is very curved near its minimum, and the field $\phi$ is strongly captured there. When the Hubble constant decreases, the minimum moves, {\it and  it drags  with it the scalar field}. As a result, the field $\phi$ almost adiabatically moves to its new equilibrium value, and the amplitude of its oscillations about it is very small. This effect reduces  strongly the energy density stored in the  oscillations of the field $\phi$ and relaxes the cosmological moduli problem.

To verify this statement, we will consider here two toy models which illustrate possible behavior of the effective potential of the field $\phi$ as a function of $H$. The simplest model was considered in \cite{LS}:
\begin{equation}
V = \frac{1}{2} m_\phi^2 \phi^2
	+ \frac{C^2}{2} H^2 \left( \phi - \phi_0 \right)^2 \ .
\label{dmod}
\end{equation}
At large $H$ the minimum appears at $\phi = \phi_0$; at small $H$ the minimum is at $\phi = 0$. Thus one would expect that the field should oscillate about $\phi = 0$ with an initial amplitude approximately equal to $\phi_0$.  The equation
of motion of the field $\phi$ in this potential is
\begin{equation} 
\label{modeom}
\ddot{\phi} + 3 H \dot{\phi} + m_\phi^2\phi
+ C^2 H^2 \left(\phi - \phi_0 \right) = 0 \ .
\end{equation}
Following \cite{LS}, we will consider the regime $H=p/t$ with $p=1/2$ for radiation domination and $p=2/3$ for
matter domination.
In the beginning  (for $ H \gg m_\phi $) $ \delta\phi = \phi_0$,
and so one can take $ \delta\phi(0) = \phi_0$ and $\dot{\delta\phi}(0)=0$.
According to \cite{LS},    the corresponding solution of Eq.~(\ref{modeom})   at large  $t \gg m_\phi^{-1}$  looks as follows:
\begin{equation}\label{solution} 
\phi \sim \frac{C^2 \phi_0}{\sqrt{\pi}}
\left( \frac{p}{2} \right)^{\frac{4-3p}{2}}
\Gamma \left( \frac{1+ \mu}{2} + \frac{\nu}{2} \right)
\Gamma \left( \frac{1+ \mu}{2} - \frac{\nu}{2} \right)
\left( \frac{H}{m_\phi} \right)^{\frac{3p}{2}}
\sin \left( m_\phi t + \frac{(2-3p)\pi}{4} \right) \ ,
\end{equation}
where $ \mu = -3(1-p)/2 $ and
$ \nu^2 = - C^2 p^2 + (3p-1)^2 /4 $.
 
It was noted in \cite{LS} that this solution  has a rather weak dependence on $C^2$ and $p$, and for  $ p = 1/2 $ or $2/3$ one has to a good approximation
\begin{equation} \label{smallC}
\phi \sim \frac{4}{3} \phi_0
\left( \frac{p}{m_\phi t} \right)^{\frac{3p}{2}}
\sin \left( m_\phi t + \frac{(2-3p)\pi}{4} \right) \ .
\end{equation}
Thus, as one could expect, the field $\phi$ oscillates  with the amplitude proportional to $\phi_0$, the factor $\left( \frac{p}{m_\phi t} \right)^{\frac{3p}{2}}$ taking care of the decrease of the initial amplitude due to the expansion of the universe.  The behavior of the field $\phi$ for the case $C \sim 1$ is illustrated by Fig. 1 a.

However, in fact the solution (\ref{solution}) has a weak dependence on $C$ only for $C \sim 1$. Meanwhile, if one takes $C \gg 1$, the behavior of the solution changes dramatically, see Fig. 1b. The field $\phi$ follows the position of the time-dependent minimum of the effective potential, and its oscillations about this position are rather small. To see these oscillations  more clearly, one should subtract from the actual value of the field $\phi$ its slowly changing mean value $\bar\phi(t)$ corresponding to the position of the time-dependent minimum of the effective potential. The result of this subtraction is shown on Fig. 2, simultaneously with the   solution (\ref{solution}), which  has the following asymptotic form\footnote{ I am very grateful to Ewan Stewart for the discussion of this asymptotic form of their solution (\ref{solution}) at large $C$. }
  for large $C$:
\begin{equation} \label{solution2}
\phi \sim \sqrt{2 p \pi}\ \phi_0\   C^{ 3p+1 \over2}\ \exp\left(-\,{C \pi p\over2 }\right)\ \left({p\over tm_\phi}\right)^{3p\over 2} \sin \left(m_\phi t+{(2-3p)\pi\over 4}\right) \ .
\end{equation}
Fig. 2 shows   numerical solution and the  analytical solution (\ref{solution2}) being superimposed. It is clearly seen that both functions coincide at large $t$, which serves as an independent verification of the validity of numerical and analytical results.

The solution (\ref{solution2}) has an amplitude  which is smaller than the amplitude of the solution (\ref{smallC}) for  $C\sim 1$ by the factor
\begin{equation} \label{solution3}
{3  \sqrt{2 p \pi}\over 4}   C^{ 3p+1 \over2}\ \exp\left(-\,{C \pi p\over2 }\right)\  .
\end{equation}
To reduce the   amplitude of oscillations, say, by the factor $10^{-10}$, which would be sufficient to solve the cosmological moduli problem, one needs $C \sim 30$  for $p = 1/2$. For $p = 2/3$ (universe dominated by nonrelativistic matter) it would be enough to have $C \sim 20$. Whereas this may look as a rather tough requirement, we remind that we do not really know the true value of this parameter.

The situation is  similar but somewhat better  for the toy model    considered in \cite{DRT}:
\begin{equation}\label{DRT} 
V=- {1\over 2}(m_{\phi}^2 + C_1^2H^2) \phi^2 + {1 \over 4 M_p^2}
(m_{\phi}^2 + C_2^2H^2) \phi^4 \ .
\label{modulimodel}
\end{equation}
For $H \gg m_{\phi}$, the minimum lies at $(C_1/C_2)M_p$.
For $H \ll m_{\phi}$, the
minimum lies at $M_p$.
The expectation was that when the Hubble parameter becomes small, the field begins to oscillate about the second minimum 
with initial amplitude  $\phi_0 \sim |1-C_1/C_2| M_p$.
This is indeed the case for $C_i \sim 1$. However, for $C_i \gg 1$ the situation is quite different.  We do not know analytical solution to the equation of motion for the theory (\ref{DRT}), but numerical solution can be easily obtained, and it shows a somewhat stronger dependence on $C_i$ than the solution for the previous model.   Let us consider, e.g., the model with $C_1 = 2C_2$. In this case the naive value of the initial amplitude of oscillations would be $M_p$. The results of a numerical investigation of the corresponding equation for the field $\phi$ shows that in order to decrease the amplitude of oscillations to the safe level of $10^{-10} M_p$ in the universe with $p = 2/3$ it would be sufficient to have, e.g., $C_1 = 12$, $C_2 = 6$. 

Thus,   the cosmological moduli problem does not appear for the fields which acquire mass approximately one order of magnitude greater than $H$ in the early universe. It is hard to tell whether this condition can be satisfied in realistic models. However,  after struggling for ten years to find a solution for the moduli problem, we should not overlook this simple possibility.  Moreover,   even if the mass of the moduli is not much greater than the Hubble constant, our results imply   that the moduli problem may be less severe than we expected, which makes the possibility of solving it by an additional short stage of inflation \cite{rt,LS,KLSphase} much more plausible.

The author  is    grateful to  M. Dine, E. Halyo,  S. Thomas, and especially to E. Stewart and D. Lyth for many valuable
discussions.  This work was
supported in part  by NSF grant PHY-8612280.

%\begin{references}

\begin{figure}
\centerline{ \epsfbox{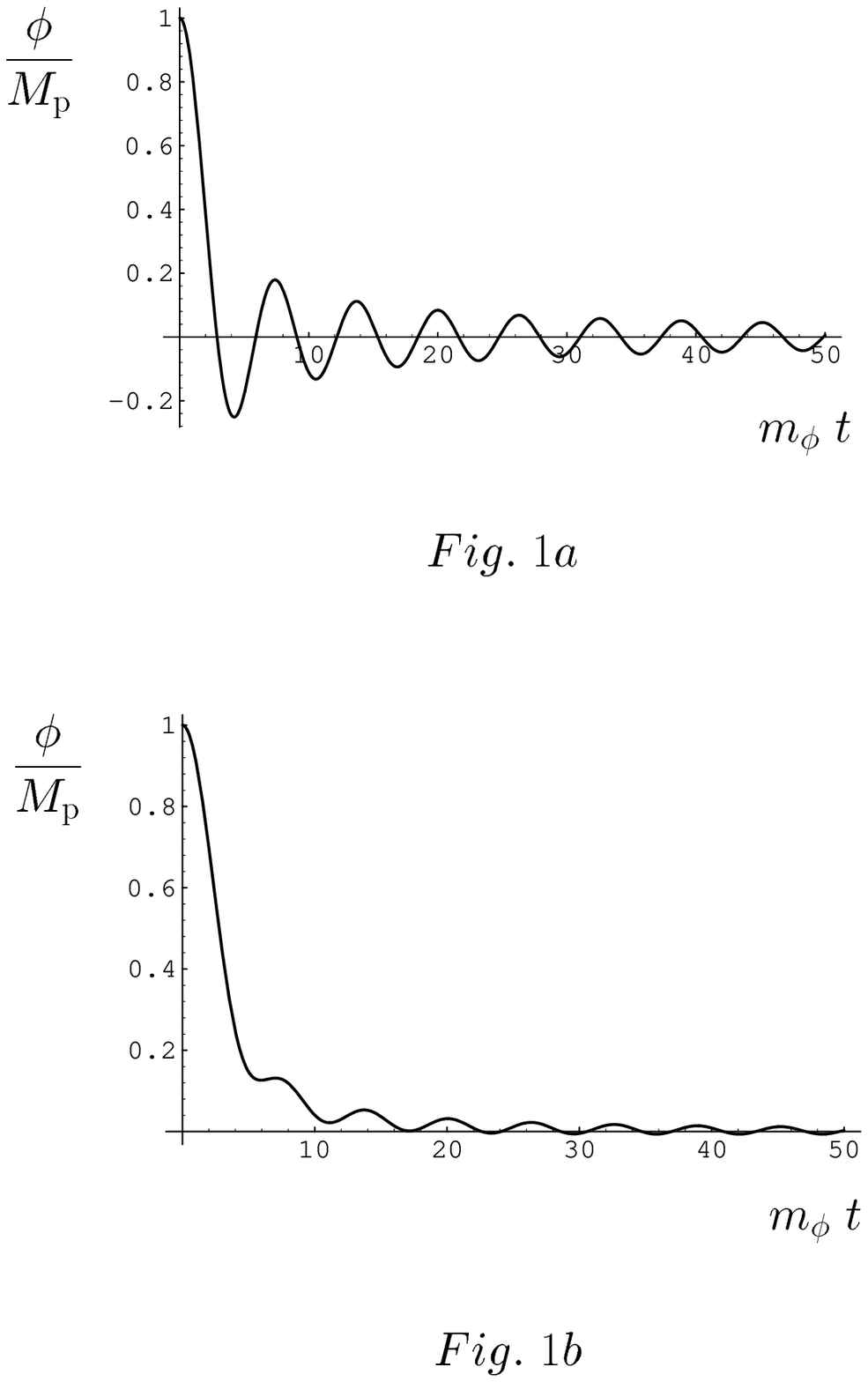}}
 \vskip 1.5cm
\caption{Oscillations of the moduli field in the theory (2) in the radiation dominated universe ($p = 1/2$). Fig. 1a corresponds to $C = 1$, Fig. 1b shows the same process for $C = 5$.}

\label{F1}

\end{figure}

\begin{figure}

\centerline{ \epsfbox{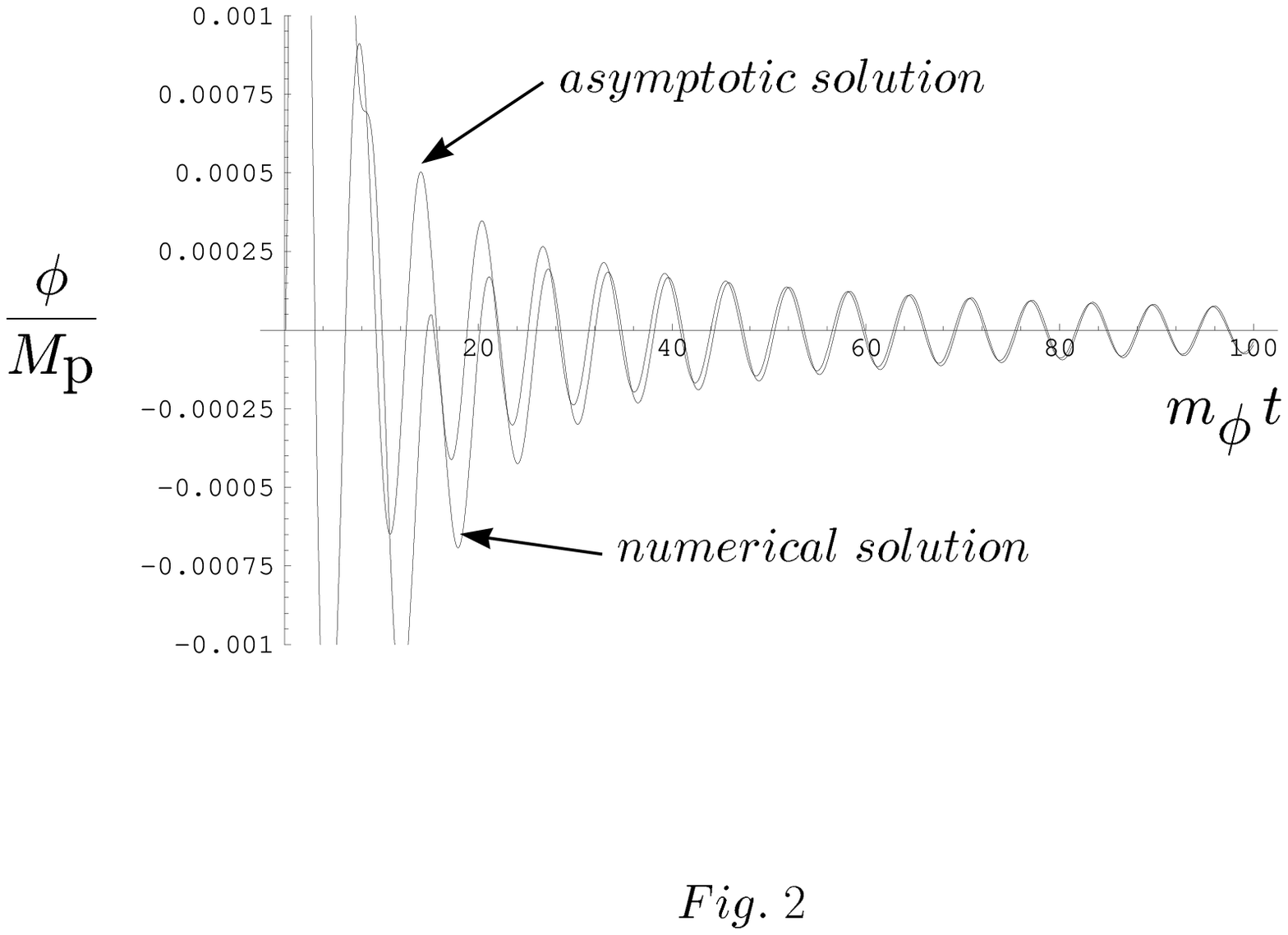}}
 \vskip 2 cm
\caption{Asymptotic solution (6) for $\phi (t)$ versus numerical solution for the deviation of the field $\phi(t)$ from the instantaneous position $\bar\phi(t)$ of the minimum of the effective potential. For definiteness, we take here $C = 8$. As expected, these two functions coincide for large $t$.}

\label{F2}

\end{figure}


\begin{thebibliography}{999} 
\bibitem{constraint} J. Ellis, G.B. Gelmini, J.L. Lopez, D.V. Nanopoulos, and S. Sarkar, Nucl. Phys. {\bf B373}, 399 (1992).

\bibitem{rt} L. Randall and S. Thomas, Nucl. Phys. {\bf B449}, 229 (1995).

\bibitem{bkn}
R. de Carlos, D.V. Nanopoulos, and
M. Quiros, Phys. Lett. B {\bf 318},  447 (1993); B. de Carlos, 
T. Banks, D. Kaplan, and A. Nelson, Phys. Rev. D {\bf 49}
(1994) 779;
T. Banks, M. Berkooz, and P. J. Steinhardt,
        Phys. Rev. D{\bf 52}, 705 (1995); G.G. Ross and S. Sarkar, ``Successful Supersymmetric Inflation,'' preprint CERN-TH.95/134 (1995), hep-ph/9506283.

\bibitem{polonyi} G. Coughlan, W. Fischler, E. Kolb, S. Raby,
and G. Ross, Phys. Lett. B {\bf 131} (1983) 59.

\bibitem{NS} D.V. Nanopoulos and M. Srednicki, Phys. Lett. {\bf 133B}, 287 (1983).

\bibitem{dinefisch} M. Dine, W. Fischler, and D. Nemeschansky,
        Phys. Lett. {\bf 136B}, 169 (1984); G. D. Coughlan, R. Holman, P. Ramond, and G. G. Ross,
        Phys. Lett. {\bf 140B}, 44 (1984).

\bibitem{Dvali} G. Dvali, ``Inflation Versus the Cosmological Moduli Problem,'' Pisa University preprint IFUP-TH-09-95 (1995), hep-ph/9503259.


\bibitem{GLV} A.S. Goncharov, A.D. Linde, and M.I. Vysotsky, Phys. Lett. {\bf 147B}, 279 (1984).

\bibitem{adshort} M. Dine, L. Randall, and S. Thomas,
Phys. Rev. Lett. {\bf 75} (1995) 398.

\bibitem{ENQ} J. Ellis,  D.V. Nanopoulos, and M. Quiros,  Phys. Lett. {\bf B174}, 176 (1984); M. Kawasaki, T. Moroi, and T. Yanagida, ``Constraint on Reheating Temperature from the Decay of the Polonyi Field,'' ICRR-340-95-6 (1995), 
  hep-ph/9509399; T. Moroi, ``On the Solution of the Polonyi Problem with No-Scale Type Supergravity,''  preprint LBL-37911 (1995), hep-ph/9510411.


\bibitem{LS}  D.H. Lyth and  E.D. Stewart, Phys. Rev. Lett. {\bf 75},
201 (1995);  D.H. Lyth and  E.D. Stewart, ``Thermal Inflation and the Moduli Problem,'' preprint  LANCASTER-TH/9505 (1995), hep-ph/9510204.

\bibitem{KLSphase} L. Kofman, A.D. Linde, and A.A. Starobinsky, ``Nonthermal Phase Transitions After Inflation,'' Stanford University preprint SU--ITP--95--21 (1995), hep-th/9510119, to be published in Phys. Rev. Lett.


\bibitem{DRT} M. Dine, L. Randall, and S. Thomas, ``Baryogenesis from Flat Directions of the Supersymmetric Standard Model,'' SLAC preprint SLAC-PUB-95-6846 (1995), hep-ph/9507453 

 \bibitem{MyBook} A.D. Linde, {\it Particle Physics
and Inflationary Cosmology} (Harwood Academic Press, Chur,
Switzerland, 1990).  

\bibitem{hybrid} A.D. Linde, Phys. Lett. {\bf B259}, 38 (1991); A.D. Linde, Phys. Rev. D {\bf 49}, 748 (1994); E. J. Copeland, A. R. Liddle, D. H. Lyth, E. D. Stewart,
        and D. Wands, Phys. Rev. D{\bf 49}, 6410 (1994).




 \end{thebibliography}
\end{document}